\documentclass[fleqn,10pt]{SelfArx} 

\usepackage{lipsum} 
\usepackage[super,comma]{natbib}

\definecolor{color1}{RGB}{0,0,90} 
\definecolor{color2}{RGB}{0,20,20} 

\usepackage{hyperref} 
\hypersetup{hidelinks,colorlinks,breaklinks=true,urlcolor=color2,citecolor=color1,linkcolor=color1,bookmarksopen=false,pdftitle={Title},pdfauthor={Author}}
\JournalInfo{Pre-print version of manuscript published at PLOS Currents Outbreaks (2 September 2014)} 
\Archive{} 

\PaperTitle{Estimating the reproduction number of Ebola virus (EBOV) during the 2014 outbreak in West Africa} 

\Authors{Christian L.~Althaus\textsuperscript{1}*} 
\affiliation{\textsuperscript{1}\textit{Institute of Social and Preventive Medicine (ISPM), University of Bern, Bern, Switzerland}} 
\affiliation{*\textbf{Corresponding author}: christian.althaus@alumni.ethz.ch} 

\newcommand{\keywordname}{Keywords} 

\Abstract{The 2014 Ebola virus (EBOV) outbreak in West Africa is the largest outbreak of the genus \emph{Ebolavirus} to date. To better understand the spread of infection in the affected countries, it is crucial to know the number of secondary cases generated by an infected index case in the absence and presence of control measures, i.e., the basic and effective reproduction number. In this study, I describe the EBOV epidemic using an SEIR (susceptible-exposed-infectious-recovered) model and fit the model to the most recent reported data of infected cases and deaths in Guinea, Sierra Leone and Liberia. The maximum likelihood estimates of the basic reproduction number are 1.51 (95\% confidence interval [CI]: 1.50-1.52) for Guinea, 2.53 (95\% CI: 2.41-2.67) for Sierra Leone and 1.59 (95\% CI: 1.57-1.60) for Liberia. The model indicates that in Guinea and Sierra Leone the effective reproduction number might have dropped to around unity by the end of May and July 2014, respectively. In Liberia, however, the model estimates no decline in the effective reproduction number by end-August 2014. This suggests that control efforts in Liberia need to be improved substantially in order to stop the current outbreak.}

\begin{document}

\flushbottom 

\maketitle 

\thispagestyle{empty} 


\section*{Introduction} 

\addcontentsline{toc}{section}{Introduction} 

The 2014 Ebola virus (EBOV) outbreak in West Africa is the largest outbreak of the genus \emph{Ebolavirus} to date. The outbreak began in Guinea in December 2013\cite{Baize:2014} and later spread to Sierra Leone, Liberia and Nigeria. While public health interventions have been introduced in all affected countries, the numbers of infected cases and deaths from EBOV continue to increase and the effects of control measures remain to be determined. Real-time analysis of the numbers of infected cases and deaths due to EBOV could provide helpful information for public health policy.

Two key parameters describing the spread of an infection are the basic and the effective reproduction numbers, $R_0$ and $R_e$, which are defined as the number of secondary infections generated by an infected index case in the absence and presence of control interventions. If $R_e$ drops below unity, the epidemic eventually stops. Several studies have fitted mathematical models to data from previous outbreaks of the genus \emph{Ebolavirus}.\cite{Chowell:2004,Lekone:2006,Legrand:2007} Previous estimates of $R_0$ from two outbreaks in Congo (1995) and Uganda (2000) range from 1.3\cite{Chowell:2004} to 2.7.\cite{Legrand:2007} It will be important to know the reproduction numbers of the current EBOV outbreak and how it is affected by public health interventions. This will facilitate making projections of the epidemic during the next months and will allow comparisons of the effects of control measures in each country.

In this study, I describe the 2014 EBOV epidemic using an SEIR (susceptible-exposed-infectious-recovered) model. Fitting the model to the most recent data about reported cases and deaths in Guinea, Sierra Leone and Liberia provided estimates of the reproduction numbers of EBOV in absence and presence of control interventions.

\begin{figure*}[ht]\centering 
\includegraphics[width=\linewidth]{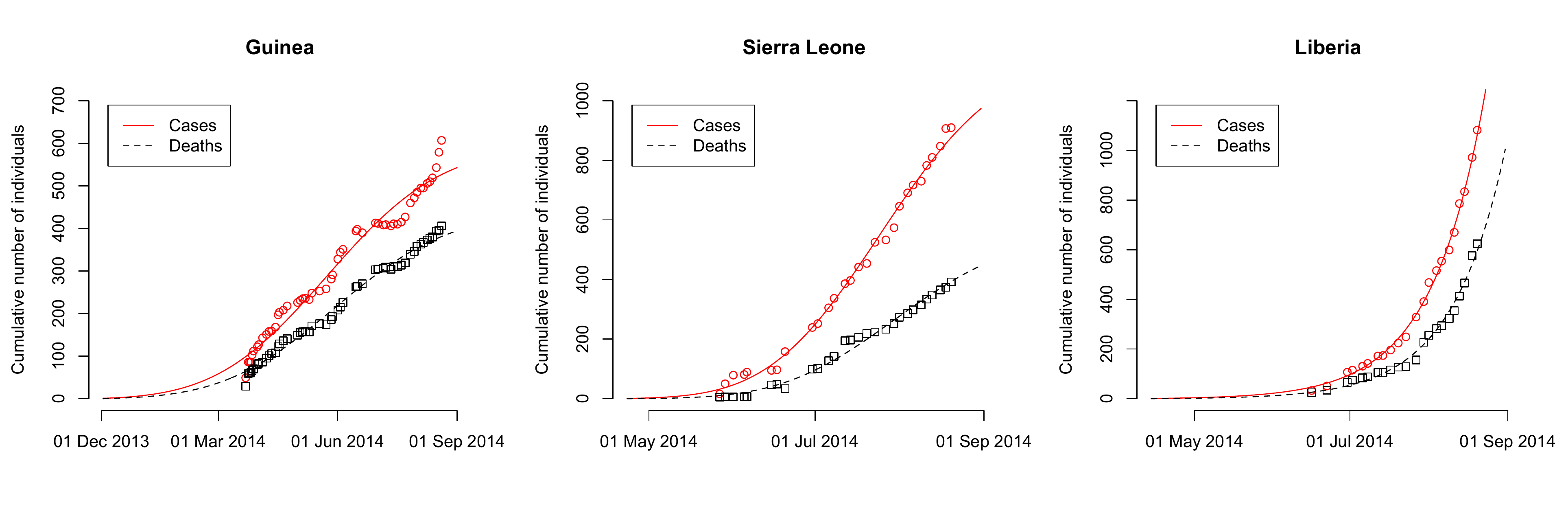}
\caption{\textbf{Dynamics of 2014 EBOV outbreaks in Guinea, Sierra Leone and Liberia.} Data of the cumulative numbers of infected cases and deaths are shown as red circles and black squares, respectively. The lines represent the best-fit model to the data. Note that the scale of the axes differ between countries.}
\label{fig:dynamics}
\end{figure*}

\begin{table*}[ht]
\caption{\textbf{Parameter estimates for the 2014 EBOV outbreak.} The basic reproduction number is given by $R_0 = \beta/\gamma$ where $1/\gamma = 5.61$ days is the infectious duration from the study by Chowell et al.\cite{Chowell:2004} The primary case is defined as the index patient that caused the subsequent outbreak. The date of appearance of the primary case in Guinea was set at 2 December 2013.\cite{Baize:2014} 95\% confidence intervals (CI) are shown in brackets. *A Likelihood ratio test showed that treating \emph{k} as a free parameter does not improve the fit.}
\centering
\begin{tabular}{lccc}
\toprule
Parameter & Guinea & Sierra Leone & Liberia \\
\midrule
Basic reproduction number, $R_0$ 		& 1.51 (1.50-1.52) 		& 2.53 (2.41-2.67)	& 1.59 (1.57-1.60)	\\
Transmission rate, $\beta$ (per day)	& 0.27 (0.27-0.27)		& 0.45 (0.43-0.48)	& 0.28 (0.28-0.29)	\\
Case fatality rate, $f$ 				& 0.74 (0.72-0.75)		& 0.48 (0.47-0.50)	& 0.71 (0.69-0.74)	\\
Rate at which control measures	 		& 0.0023				& 0.0097			& 0*				\\
reduce transmission, $k$ (per day) 		& (0.0023-0.0024)		& (0.0085-0.0110)	&					\\
Date of appearance of primary case, $T$ & -- 					& 23 Apr 2014		& 14 April 2014 	\\
										&						& (19-25 Apr 2014)	& (11-16 Apr 2014)	\\
\bottomrule
\end{tabular}
\label{tab:estimates}
\end{table*}

\begin{figure*}[ht]\centering 
\includegraphics[width=\linewidth]{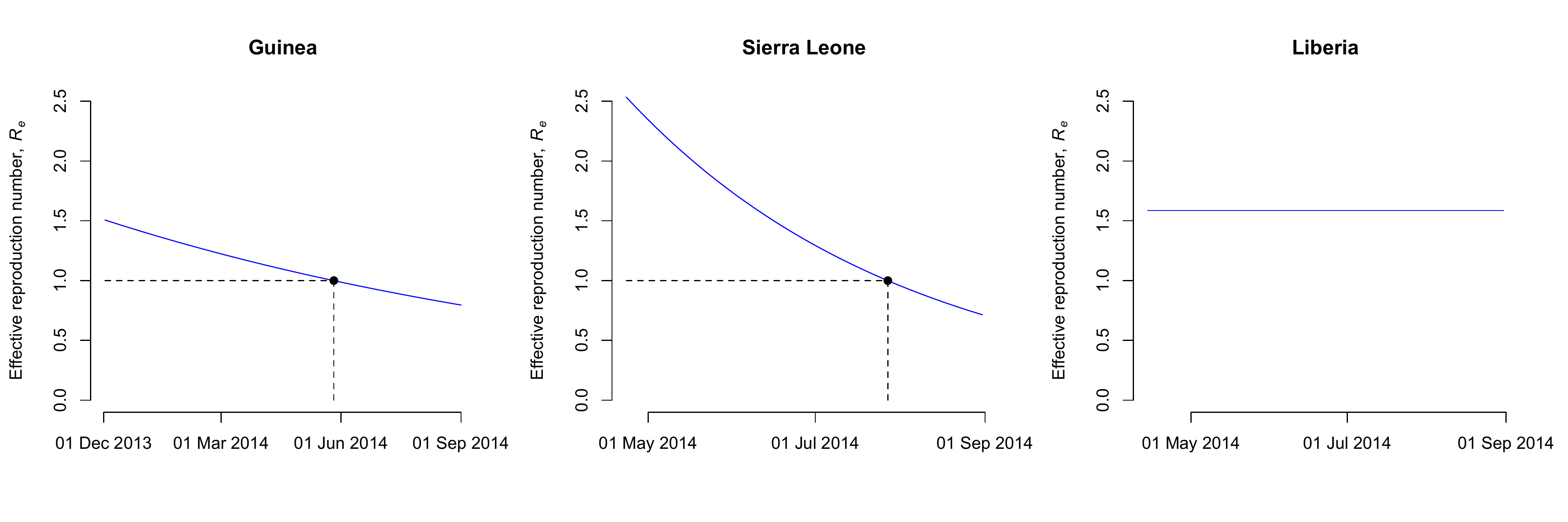}
\caption{\textbf{Effective reproduction number of EBOV in Guinea, Sierra Leone and Liberia.} The model assumes that the transmission rate decays exponentially due to the introduction of control measures.  In Guinea and Sierra Leone, the effective reproduction number has dropped to around unity by the end of May and July 2014, respectively (dashed lines). In Liberia, the effective reproduction number remains unchanged by end-August 2014. Note that the scale of the x-axis differs between countries.}
\label{fig:Re}
\end{figure*}

\section*{Methods}
The transmission of EBOV follows SEIR (susceptible-exposed-infectious-recovered) dynamics and can be described by the following set of ordinary differential equations (ODEs):\cite{Chowell:2004}
\begin{align}
\frac{dS}{dt} &= -\beta(t)SI/N, \\
\frac{dE}{dt} &= \beta(t)SI/N - \sigma E, \\
\frac{dI}{dt} &= \sigma E - \gamma I, \\
\frac{dR}{dt} &= (1-f)\gamma I.
\end{align}

After transmission of the virus, susceptible individuals $S$ enter the exposed class $E$ before they become infectious individuals $I$ that either recover and survive ($R$) or die. $1/\sigma$ and $1/\gamma$ are the average durations of incubation and infectiousness. The case fatality rate is given by $f$. The transmission rate in absence of control interventions is constant, i.e., $\beta(t) = \beta$. After control measures are introduced at time $\tau \leq t$, the transmission rate was assumed to decay exponentially at rate $k$:\cite{Lekone:2006}
\begin{align}
\beta(t) &= \beta e^{-k(t-\tau)},
\end{align}
i.e., the time until the transmission rate is at 50\% of its initial level is $t_{1/2} = ln(2)/k$. Assuming the epidemics starts with a single infected case ($I_0 = 1$ and $C_0 = 1$), the cumulative number of infected cases $C$ and deaths $D$ are given by the solutions to $dC/dt = \sigma E$ and $dD/dt = f \gamma I$, respectively. The ODEs were solved numerically in the R software environment for statistical computing\cite{R:2014} using the function \emph{ode} from the package \emph{deSolve}. 

Outbreak data for Guinea, Sierra Leone and Liberia (Table \ref{tab:data}) were based on the cumulative numbers of reported total cases (confirmed, probable and suspected) and deaths from the World Health Organisation (WHO).\cite{Ebola:2014} The total population size $N = S + E + I + R$ in each country was assumed to be $10^6$ individuals. Note that the exact population size does not need to be known to estimate the model parameters as long as the number of cases is small compared to the total population size.\cite{Chowell:2004} In particular, the basic reproduction number is simply given by $R_0 = \beta/\gamma$. The effective reproduction number is given by $R_e = \beta(t)S/(\gamma N) \approx \beta(t)/\gamma$ as long as the number of cases remains much smaller than $S$.

Maximum likelihood estimates of the parameters were obtained by fitting the model to the data, assuming the cumulative numbers of cases and deaths are Poisson distributed. I used the optimization algorithm by Nelder \& Mead that is implemented in the function \emph{optim}. 95\% confidence intervals (CI) were calculated from the likelihood profile.

\section*{Results}
The limited number of data items and the fact that the EBOV outbreak is ongoing prevent the estimation of all model parameters. Hence, I made three simplifying assumptions. First, the average duration of the incubation and infectious period were fixed to previous estimates from an outbreak of the same EBOV subtype in Congo in 1995 ($1/\sigma = 5.3$ days and $1/\gamma = 5.61$ days).\cite{Chowell:2004} Second, I assumed that control measures began after the appearance of the first infected case, i.e., $\tau = 0$. Third, the date of appearance of the primary case of the outbreak in Guinea was set to 2 December 2013.\cite{Baize:2014} The remaining parameters to be estimated are the transmission rate $\beta$, the case fatality rate $f$, the rate $k$ at which control measures reduce the transmission rate and the time of appearance of the primary case that caused the subsequent outbreak $T$ (for Sierra Leone and Liberia only).

The model fits the reported data of cases and deaths in Guinea, Sierra Leone and Liberia well (Figure \ref{fig:dynamics}). The maximum likelihood estimates of the basic reproduction number, $R_0$, are 1.51 (95\% CI: 1.50-1.52) for Guinea, 2.53 (95\% CI: 2.41-2.67) for Sierra Leone and 1.59 (95\% CI: 1.57-1.60) for Liberia (Table \ref{tab:estimates}). The case fatality rate \emph{f} is estimated at 74\% (95\% CI: 72\%-75\%) for Guinea, 48\% (95\% CI: 47\%-50\%) for Sierra Leone and 71\% (95\% CI: 69\%-74\%) for Liberia. The estimates of the parameter \emph{k}, which describes how the control measures reduce the transmission rate, vary between countries (Table \ref{tab:estimates}). This results in a different decrease of the effective reproduction number, $R_e$, after the outbreaks started in each country. While $R_e$ seems to have dropped to levels around unity in Guinea and Sierra Leone by end-August 2014, the model suggests that control interventions were not successful in reducing $R_e$ in Liberia (Fig.~\ref{fig:Re}).

\section*{Discussion}
This study uses mathematical modeling to estimate the basic and effective reproduction numbers of EBOV during the 2014 outbreak in West Africa. The maximum likelihood esitmates of $R_0$ are 1.51 for Guinea, 2.53 for Sierra Leone and 1.59 for Liberia and lie within the same range as previous estimates for an EBOV outbreak in Congo (1995) and an outbreak of Sudan virus (SUDV) in Uganda (2000).\cite{Chowell:2004,Lekone:2006,Legrand:2007} The basic reproduction number in Sierra Leone seems to be significantly higher than in Guinea and Liberia. This could be a result of differences in the population structure or human mobility.

This study provides real-time estimates of EBOV transmission parameters during an ongoing outbreak. The mathematical model of transmission is informed by previous studies\cite{Chowell:2004,Lekone:2006,Legrand:2007} and provides a good description of the current epidemic. Interestingly, the estimated date of appearance of the primary case that caused the subsequent outbreak in Sierra Leone is in agreement with the time of the most recent common ancestor of a genome analysis of the EBOV outbreak.\cite{Gire:2014}

The simplifying assumptions of the model mean that the results of this study need to be interpreted with caution. A major limitation of the model is that the transmission rate decays exponentially due to control measures after the appearance of the first infectious case. Thus, the model cannot account for fluctuations in the number of new cases as seen in Guinea. These fluctuations could be a result of varying effects of control interventions in different areas, suggesting that $R_e$ in some parts such as urban areas might well be above unity. As more data of the EBOV outbreak is becoming available, this modeling aspect might fit better to data from specific areas within a country. The effectiveness of current control measures on the transmission rate is unknown at present and the size and duration of the outbreak suggest that they need to improve. As more data accumulate over the next weeks and months a more thorough analysis will allow more accurate estimates of $R_0$ and $R_e$ together with epidemic projections and the uncertainty around them.

Real-time estimates of epidemic parameters and model projections are important for predicting the evolution of this EBOV epidemic and investigating the effects of ongoing and new interventions. Hygiene measures and social distancing intervention are already being implemented and WHO has given permission for the use and evaluation of currently unregistered new drugs and vaccines. The results of this study suggest that control measures might have been sufficient to decrease the effective reproduction number to around unity in Guinea and Sierra Leone by the end of May and July 2014, respectively. In Liberia, however, control efforts still need to be improved substantially in order to stop the current outbreak.

\phantomsection
\section*{Acknowledgments} 
I would like to thank Nicola Low and Sandro Gsteiger for helpful comments on the manuscript.

\section*{Funding Statement} 
Christian L.~Althaus is funded by an Ambizione grant from the Swiss National Science Foundation (project 136737). The funders had no role in study design, data collection and analysis, decision to publish, or preparation of the manuscript.

\section*{Competing Interests} 
The author has declared that no competing interests exist.

\addcontentsline{toc}{section}{Acknowledgments} 

\phantomsection

\begin{table*}[ht!]
\caption{\textbf{Reported cumulative numbers of cases and deaths of EBOV during the 2014 outbreak.} Data were retrieved from the WHO website and based on the cumulative total numbers of clinical cases (confirmed, probable and suspected).\cite{Ebola:2014} The first confirmed cases in Sierra Leone were reported on 27 May 2014. In Liberia, the cases that were reported on 16 June 2014 were the first new cases since 6 April 2014.}
\centering
\footnotesize
\begin{tabular}{crrrrrr}
\toprule
\textbf{Date of report}	&	\multicolumn{2}{c}{\textbf{Guinea}}		&	\multicolumn{2}{c}{\textbf{Sierra Leone}}		&	\multicolumn{2}{c}{\textbf{Liberia}}	\\
						&	\textbf{Cases}	&	\textbf{Deaths}		&	\textbf{Cases}	&	\textbf{Deaths}				&	\textbf{Cases}	&	\textbf{Deaths}		\\
\midrule
22 Mar 2014	&	49	&	29	&		&		&		&		\\
24 Mar 2014	&	86	&	59	&		&		&		&		\\
25 Mar 2014	&	86	&	60	&		&		&		&		\\
26 Mar 2014	&	86	&	62	&		&		&		&		\\
27 Mar 2014	&	103	&	66	&		&		&		&		\\
28 Mar 2014	&	112	&	70	&		&		&		&		\\
31 Mar 2014	&	122	&	80	&		&		&		&		\\
01 Apr 2014	&	127	&	83	&		&		&		&		\\
04 Apr 2014	&	143	&	86	&		&		&		&		\\
07 Apr 2014	&	151	&	95	&		&		&		&		\\
09 Apr 2014	&	158	&	101	&		&		&		&		\\
11 Apr 2014	&	159	&	106	&		&		&		&		\\
14 Apr 2014	&	168	&	108	&		&		&		&		\\
16 Apr 2014	&	197	&	122	&		&		&		&		\\
17 Apr 2014	&	203	&	129	&		&		&		&		\\
20 Apr 2014	&	208	&	136	&		&		&		&		\\
23 Apr 2014	&	218	&	141	&		&		&		&		\\
01 May 2014	&	226	&	149	&		&		&		&		\\
03 May 2014	&	231	&	155	&		&		&		&		\\
05 May 2014	&	235	&	157	&		&		&		&		\\
07 May 2014	&	236	&	158	&		&		&		&		\\
10 May 2014	&	233	&	157	&		&		&		&		\\
12 May 2014	&	248	&	171	&		&		&		&		\\
18 May 2014	&	253	&	176	&		&		&		&		\\
23 May 2014	&	258	&	174	&		&		&		&		\\
27 May 2014	&	281	&	186	&	16	&	5	&		&		\\
28 May 2014	&	291	&	193	&		&		&		&		\\
29 May 2014	&		&		&	50	&	6	&		&		\\
01 Jun 2014	&	328	&	208	&	79	&	6	&		&		\\
03 Jun 2014	&	344	&	215	&		&		&		&		\\
05 Jun 2014	&	351	&	226	&	81	&	6	&		&		\\
06 Jun 2014	&		&		&	89	&	7	&		&		\\
15 Jun 2014	&	394	&	263	&	95	&	46	&		&		\\
16 Jun 2014	&	398	&	264	&		&		&	33	&	24	\\
17 Jun 2014	&		&		&	97	&	49	&		&		\\
20 Jun 2014	&	390	&	270	&	158	&	34	&		&		\\
22 Jun 2014	&		&		&		&		&	51	&	34	\\
30 Jun 2014	&	413	&	303	&	239	&	99	&	107	&	65	\\
02 Jul 2014	&	412	&	305	&	252	&	101	&	115	&	75	\\
06 Jul 2014	&	408	&	307	&	305	&	127	&	131	&	84	\\
08 Jul 2014	&	409	&	309	&	337	&	142	&	142	&	88	\\
12 Jul 2014	&	406	&	304	&	386	&	194	&	172	&	105	\\
14 Jul 2014	&	411	&	310	&	397	&	197	&	174	&	106	\\
17 Jul 2014	&	410	&	310	&	442	&	206	&	196	&	116	\\
20 Jul 2014	&	415	&	314	&	454	&	219	&	224	&	127	\\
23 Jul 2014	&	427	&	319	&	525	&	224	&	249	&	129	\\
27 Jul 2014	&	460	&	339	&	533	&	233	&	329	&	156	\\
30 Jul 2014	&	472	&	346	&	574	&	252	&	391	&	227	\\
01 Aug 2014	&	485	&	358	&	646	&	273	&	468	&	255	\\
04 Aug 2014	&	495	&	363	&	691	&	286	&	516	&	282	\\
06 Aug 2014	&	495	&	367	&	717	&	298	&	554	&	294	\\
09 Aug 2014	&	506	&	373	&	730	&	315	&	599	&	323	\\
11 Aug 2014	&	510	&	377	&	783	&	334	&	670	&	355	\\
13 Aug 2014	&	519	&	380	&	810	&	348	&	786	&	413	\\
15 Aug 2014	&		&		&		&		&	834	&	466	\\
16 Aug 2014	&	543	&	394	&	848	&	365	&		&		\\
18 Aug 2014	&	579	&	396	&	907	&	374	&	972	&	576	\\
20 Aug 2014	&	607	&	406	&	910	&	392	&	1082	&	624	\\
\bottomrule
\end{tabular}
\normalsize
\label{tab:data}
\end{table*}



\begin{thebibliography}{6}
\providecommand{\natexlab}[1]{#1}
\expandafter\ifx\csname urlstyle\endcsname\relax
  \providecommand{\doi}[1]{doi:\discretionary{}{}{}#1}\else
  \providecommand{\doi}{doi:\discretionary{}{}{}\begingroup
  \urlstyle{rm}\Url}\fi

\bibitem[{Baize \emph{et~al.}(2014)Baize, Pannetier, Oestereich, Rieger,
  Koivogui, Magassouba, Soropogui, Sow, Ke{\"\i}ta \emph{et~al.}}]{Baize:2014}
Baize, S., Pannetier, D., Oestereich, L., Rieger, T., Koivogui, L., Magassouba,
  N., Soropogui, B., Sow, M.~S., Ke{\"\i}ta, S. \emph{et~al.} 2014 {Emergence
  of Zaire Ebola Virus Disease in Guinea - Preliminary Report}.
\newblock \emph{N Engl J Med} \doi{10.1056/NEJMoa1404505}.

\bibitem[{Chowell \emph{et~al.}(2004)Chowell, Hengartner, Castillo-Chavez,
  Fenimore \& Hyman}]{Chowell:2004}
Chowell, G., Hengartner, N.~W., Castillo-Chavez, C., Fenimore, P.~W. \& Hyman,
  J.~M. 2004 {The basic reproductive number of Ebola and the effects of public
  health measures: the cases of Congo and Uganda}.
\newblock \emph{J Theor Biol} \textbf{229}, 119--26.
\newblock \doi{10.1016/j.jtbi.2004.03.006}.

\bibitem[{Lekone \& Finkenst{\"a}dt(2006)}]{Lekone:2006}
Lekone, P.~E. \& Finkenst{\"a}dt, B.~F. 2006 {Statistical inference in a
  stochastic epidemic SEIR model with control intervention: Ebola as a case
  study}.
\newblock \emph{Biometrics} \textbf{62}, 1170--7.
\newblock \doi{10.1111/j.1541-0420.2006.00609.x}.

\bibitem[{Legrand \emph{et~al.}(2007)Legrand, Grais, Boelle, Valleron \&
  Flahault}]{Legrand:2007}
Legrand, J., Grais, R.~F., Boelle, P.~Y., Valleron, A.~J. \& Flahault, A. 2007
  {Understanding the dynamics of Ebola epidemics}.
\newblock \emph{Epidemiol Infect} \textbf{135}, 610--21.
\newblock \doi{10.1017/S0950268806007217}.

\bibitem[{Ebo()}]{Ebola:2014}
 {Disease Outbreak News -- WHO Regional Office for Africa:}
  \url{http://www.afro.who.int/en/clusters-a-programmes/dpc/epidemic-a-pandemic-alert-and-response/outbreak-news.html}.

\bibitem[{{R Development Core Team}(2014)}]{R:2014}
{R Development Core Team} 2014 \emph{{R: {A} {L}anguage and {E}nvironment for
  {S}tatistical {C}omputing}}.
\newblock R Foundation for Statistical Computing, Vienna, Austria.

\bibitem[{Gire \emph{et~al.}(2014)Gire, Goba, Andersen, Sealfon,
  Park, Kanneh, Jalloh, Momoh, Fullah \emph{et~al.}}]{Gire:2014}
Gire, S.~K., Goba, A., Andersen, K.~G., Sealfon, R.~S., Park, D.~J., Kanneh,
  L., Jalloh, S., Momoh, M., Fullah, M. \emph{et~al.} 2014 {Genomic surveillance elucidates Ebola virus origin and transmission during the 2014 outbreak}.
\newblock \emph{Science} \doi{10.1126/science.1259657}.

\end{thebibliography}
\end{document}